\documentstyle[12pt,aasms4]{article}
\lefthead{Olech et al.}
\righthead{RR Lyr Variables in M55}
 
\begin{document}
 
\title{RR Lyrae Variables in the Globular Cluster M55. \\
The First
Evidence for Non Radial Pulsations in RR Lyr Stars.}
\author{A. Olech\altaffilmark{1}, ~J. Kaluzny\altaffilmark{1}, ~I. ~B.
Thompson\altaffilmark{2}, 
~W. Pych\altaffilmark{1},
~W. Krzeminski\altaffilmark{2,4} \and ~A.
Schwarzenberg-Czerny\altaffilmark{3,4}}
 
\altaffiltext{1}{Warsaw University Observatory, Al. Ujazdowskie 4,
00-478
Warsaw, Poland (olech,jka,pych@sirius.astrouw.edu.pl)}
\altaffiltext{2}
{Carnegie Institution of Washington, 813 Santa Barbara Street, Pasadena,
CA 91101, USA (ian@ociw.edu, wojtek@lco.cl)}
\altaffiltext{3}
{Astronomical Observatory of Adam Mickiewicz University,
ul. Sloneczna 36, Poznan, Poland}
\altaffiltext{4}
{Copernicus Astronomical Center,
ul. Bartycka 18, 00-716 Warsaw, Poland
(wk@camk.edu.pl,alex@camk.edu.pl)}

\begin{abstract}
 
We present the results of a photometric study of RR Lyrae variables in
the field of the globular cluster M55. We have discovered nine new RR
Lyrae stars, increasing the number of known variables in this cluster to
15 objects. Five of the newly discovered variables belong to Bailey type
RRc  and two to type RRab. Two background  RRab stars are probable
members of the Sagittarius dwarf galaxy. Fourier decomposition of the
light curves was used to derive basic properties of the present sample
of RR Lyrae variables. From an analysis of the RRc variables we obtain a
mean mass of $M=0.53\pm0.03 M_\odot$, luminosity  $\log L=1.75\pm0.01$,
effective temperature $T_{eff}=7193\pm27$~K, and helium abundance
$Y=0.27\pm0.01$.  Based on the $B-V$ colors, periods and metallicities
of the RRab stars we estimate the value of the color excess for M55 to
be equal to $E(B-V)=0.11\pm0.03$. Using this value we derive the colors
of the  blue and red edges of the instability strip in M55. The blue
edge lies at $(B-V)_0=0.20$ mag and the red edge lies at $(B-V)_0=0.38$
mag. We estimate the values of the visual apparent and dereddened
distance moduli to be $13.65\pm0.11$ and $13.31\pm0.11$, respectively. 
The light curves of three of the RRc variables exhibit changes in
amplitude of over 0.1 mag on the time scale of less than a week, rather
short for the Blazhko effect, but with no evidence for another radial
pulsational frequency. However we do  detect other periodicities which
are clearly visible in the light curve after removing variations with
the first overtone radial frequency. This is strong evidence for the
presence of non-radial pulsations, a behavior common for $\delta$ Scuti
stars but not yet observed among RR Lyr variables.
 
\end{abstract}
 

\section{Introduction}
 
M55 (=NGC~6809, =C1936-310) is a metal-poor halo globular cluster that
is rich and at the same time relatively easy to study owing to its
proximity and relative openness.  Its reddening and apparent distance
moduli are estimated at $E(B-V)=0.07 $ and $(m-M)_{V}=13.76 $  (Harris
1996).  The cluster was selected as one of the targets in an ongoing
survey for eclipsing binaries in globular clusters (eg. Kaluzny,
Thompson \& Krzeminski 1997). As a side result of that survey we
obtained extensive time-series photometry for several RR~Lyr stars
belonging to M55. More than half of these variables are new discoveries.
This contribution is devoted exclusively to the presentation and
analysis of photometry of M55 RR~Lyr variables from the cluster field. 
Results obtained for other variables will be published elsewhere
(Thompson et al., in preparation).
 
\section{Observations and Data Reduction} 
 
Time-series photometry of M55 was obtained during the interval 1997 May
09 -- September 17 with the 1.0-m Swope telescope at Las Campanas
Observatory. The CCD camera  used for the observations  has a field of
view of $14.5' \times 23'$ with a scale of 0.435 arcsec/pixel.  More
than  700 $V$-band frames and 65 $B$-band frames were obtained with
exposure times ranging from 150 to 300 seconds  for the $V$ filter and
200 to 360 seconds for the $B$ filter  depending on the seeing.
Instrumental photometry was measured using DoPHOT (Schechter, Mateo \&
Saha 1993) and the transformation to the standard $BV$ system was based
on observations of several Landolt fields (Landolt 1992).  A more
detailed description of the observations and reductions can be found in
a complementary  paper presenting results for eclipsing binaries and
SX~Phe stars identified in M55 (Thompson et al., in preparation).
 
\section{Results}
 
This search for variable stars in M55 has identified 15 RR Lyrae
variables. Six of them (stars V1--V6) were previously known (Bailey
1902, King 1951, King and Bruzual 1976). Of the remaining nine 
newly discovered variables, 5 are type RRc   and 4 are type RRab.
 
The periods, intensity averaged $V$ brightnesses, amplitudes, $<B>-<V>$
colors and equatorial coordinates of the 15 RR Lyr variables are listed
in Table 1. A color-magnitude diagram of M55 derived from our
observations of M55 is presented in Fig. 1. Open circles denote the RRc
stars and filled circles correspond to the RRab variables belonging to
M55. The filled triangles correspond to two RRab stars that are more
distant than M55 by about 3.5 magnitudes. These two variables are most
likely members of the Sagittarius dwarf galaxy (Ibata et al. 1994).
The presence of a noticeable population of stars belonging to the
Sagittarius dwarf in the field of M55 has been  noted by Mateo \&
Mirabel (1996) and Fahlman et al. (1996). Moreover the first authors
reported identification of three RR Lyr stars from the Sagittarius
dwarf in their survey for variables in M55. It is likely that the two
distant RRab stars discovered by us are among the variables already
noted by Mateo \& Mirabel (1996).  The periods of the cluster RRc stars
are between 0.310 and 0.406 days with a mean period of 0.36 days. The
periods of the cluster RRab variables are between 0.580 and 0.722 days
with a mean value of 0.66 days. These properties clearly place M55
among the Oosterhoff type II clusters.
 
We have fit our $V$-band light curves to Fourier
series with the form:
 
\begin{equation}
V = A_0 + \sum^{10}_{j=1} A_j\cdot\sin(j\omega t + \phi_j)
\end{equation}
 
\noindent where $\omega=2\pi/P$ and $P$ is the pulsation period of the
star. A method developed by Schwarzenberg-Czerny (1997) was used to
determine values of $\omega$, $A_{j}$ and $\phi_{j}$. Although the
formal errors of our periods from a least squares  fit are quite small,
we estimate their actual value to be as large as 0.000010. This is
caused by correlations of residuals not accounted for in the least
squares solution. Our estimate for the errors in the derived periods
corresponds to  a phase uncertainty of $0.01\cdot P$ over the entire
length of the observations, a reasonable value given the quality of our
light curves. Its magnitude is in agreement with the scatter of periods
obtained by fitting Fourier series with  different numbers of
harmonics.
 
In Figs. 2, 3 and 4 we show $V$, $B$ and $B-V$ light curves of the
M55 RRab stars, the M55  RRc stars,  and the RRab stars from the
Sagittarius dwarf galaxy, respectively. The $B-V$ colors of each star
were derived from equation (1) by calculating the $V$ brightness at the
epoch of each of the $B$ observations.
 
The values of the peak to peak amplitudes $A_V$ presented in Table 1
are used to plot the period-amplitude diagram shown in Fig. 5. Again
open circles denote RRc stars, filled circles RRab stars, and filled
triangles the  two Sagittarius dwarf RRab stars identified in the field
of M55.  The solid line represents a linear fit to RRab variables in M3
(Kaluzny et al 1998).  The data presented in Fig. 5 agree with  the
well established fact that metallicity of M55 is lower than the
metallicity of M3. At the same time the metallicities of the two RRab
stars from the Sagittarius dwarf are likely to be slightly higher than
the metallicity of M3.
 
\subsection{A Fourier analysis of the RRc variables}
 
In a series of papers Simon and Teays (1982), Simon (1989) and Simon
and Clement (1993) have presented  a  method of estimating the masses,
luminosities, effective temperatures and the helium abundance of RRc
stars based only on a Fourier decomposition of the $V$-band light
curves.  The equations of Simon and Clement (1993) are:
 
\begin{equation}
\log M = 0.52\log P - 0.11\phi_{31} + 0.39
\end{equation}
 
\begin{equation}
\log L = 1.04\log P - 0.058\phi_{31} + 2.41
\end{equation}
 
\begin{equation}
\log T_{eff} = 3.265 - 0.3026\log P - 0.1777\log M + 0.2402\log L
\end{equation}
 
\begin{equation}
\log Y = -20.26 + 4.935\log T_{eff} - 0.2638\log M+ 0.3318\log L
\end{equation}
 
\vspace{0.1cm}
 
\noindent where $M$ is the mass of the star in solar units, $P$ is the
first overtone pulsation period in days, $L$ is the luminosity in 
solar units, $T_{eff}$ the effective temperature in Kelvins and
$\phi_{31}=\phi_3-3\phi_1$ (cf. equation 1).
 
Using the above equations we computed the masses, luminosities and
effective temperatures of the RRc stars in our sample. These are
presented in Table 2 together with the values of $A_0$, $A_1$ and
$\phi_{31}$.  The errors presented in Table 2 are calculated
from the error propagation law. Due to the
fact that Simon and Clement (1993) used a cosine Fourier series the
values of $\phi_{31}$ in Table 2 differ from their typical values by
$\pi$.
 
We exclude from our sample variables V9, V10 and V12 due to their
irregular light curves (see subsection 3.2 for a detailed discussion).
We also omit variable V11 due to the low accuracy of the estimation of
$\phi_{31}$. The mean values of the mass, luminosity, effective
temperature, and helium abundance for the remaining 5 RRc variables are
$0.53\pm0.03 M_\odot$, $\log {{L}\over{L_\odot}}=1.75\pm0.01$,
$T_{eff}$=$7193\pm27$~K, and $Y=0.27\pm0.01$, respectively.  These
values are broadly consistent with those for the RRc variables in the
sequence of globular clusters discussed  by Kaluzny et al (1998) (cf.
their Table 4). In particular we conclude that M55 is similar to NGC
2298.
The mean
$\log L$ and $T_{eff}$ for this cluster are 1.75 and 7200~K,
respectively, with a helium abundance of $Y$ = 0.26.
 
Previous determinations of the metallicity of M55 vary between $-1.54$
(Bica and Pastoriza 1983, Pilachowski 1984, Smith 1984) and $-1.89$
(Zinn 1980, Richtler 1988). The lower estimates agree  with those of
Suntzeff et al (1991) who obtained ${\rm [Fe/H]}=-1.81$ and McWilliam et
al (1992) who obtained ${\rm [Fe/H]}=-1.91$. Considering that the values
of $\log L$, $T_{eff}$ and $Y$ obtained  from the RRc stars are
consistent with the sequence from Kaluzny et al (1998), we conclude that
this analysis supports the more metal-poor determinations of [Fe/H] for
M55.
 
However, the Fourier analysis suggests a mean mass for the M55 RRc stars
that is discrepant with the Kaluzny et al (1998) sequence. Our mean
value is $0.53\pm0.03 M_\odot$. It places M55 between globular clusters
such as NGC 6171  ([Fe/H]=$-0.68$, $\log L = 1.65$ and $T_{eff}$=7447~K)
and M5 ([Fe/H]=$-1.25$, $\log L = 1.68$ and $T_{eff}$=7388~K), these
values disagree markedly with the values determined in this present
work.
 
We must take into account that the Simon, Teays and Clement formulae for
estimates of $\log M$, $\log L$, $\log T_{eff}$ and $Y$ are all linear
functions of $\log{P}$ and $\phi_{31}$. Let us assume for simplicity
that the distributions of $\log{P}$ and $\phi_{31}$ in the cluster are
independent, with standard deviations of 0.04 and 0.6, respectively.

Then by propagation of errors rule one can calculate the covariance
matrix of $\log M$, $\log L$, $\log T_{eff}$ and $Y$. The resulting
correlation coefficients are all above 0.9, except for $-0.78$ for the
correlation of $\log M$ and $\log T_{eff}$. This may suggest that the
Simon, Teays and Clement formulae do not effectively use all of the
information contained in $\log P$ and $\phi_{31}$. For this reason it
may be preferable to consider physical similarities between different
clusters directly on the $\log P$ and $\phi_{31}$ plane. For our sample
of the RRc variables in M55 these values are $-0.425\pm0.017$ and
$4.00\pm0.14$, respectively.
 
In Fig. 6 we show a plot of $\log L$ against mean $V$. The solid lines
have a slope of 0.4 and are separated by 0.04 in $\log L$, which
represents the uncertainty in the values of $\log L$ computed from
$\phi_{31}$ and $P$. It is clear that our points do not lie between
these lines. But the accuracy in the estimates for the masses, the
temperatures and the luminosities does not come from the experimental
errors in the measurements of the periods and the values of $\phi_{31}$,
but probably in the systematic errors in the Simon and Clement (1993)
analysis.

\subsection{The non radial pulsation of RRc stars.}
 
Three of the RRc variables presented in Fig. 3, namely V9, V10 and V12
display a modulation of their light curves. On certain occasions the
amplitudes of the light curves  change on the time scale of days. This
is shown in Fig. 7, where we present average light curves with selected
light curves from individual nights. Such behavior is not typical for
stars exhibiting the Blazhko effect (Blazhko 1907; see also the
discussion in Smith 1995). In particular, in a sample of 46 stars for
which Blazhko periods have been determined, the shortest period is 10.9
days. We find no statistically significant periods for V9, V10, and V12
in the range 5 days to 60 days.
 
This behavior is common in bimodal pulsators (RR Lyr type RRd). To
clarify the nature of this modulation we performed a period analysis of
the light curves.  Power spectra, CLEAN (Roberts et al. 1987) and
multi-harmonic periodograms of raw light curves reveal no radial
periods other than those listed in Table 1 and their  harmonics and $1
c/d$ aliases.  However, the periods of all 3 stars are very close to
1/3 of a day so that the same portions of the light curves are observed
over several weeks. In order to judge  the magnitude of any effects of
aliasing, we  simulated $V$-band observations of a bimodal pulsating
star with $P_1$ from Table 1 and overtone-to-fundamental period ratio
$P_1/P_0$ = 0.745, typical for RRd variables (Smith 1995), using the
formula:
 
$$V(t) = 0.05\cdot\sin{(2\pi t/P_0)}+0.22\cdot\sin{(2\pi t/P_1)}+n(t)$$ 
 
\noindent where $n(t)$ denotes a Gaussian noise component with standard
deviation 0.025 mag.  Synthetic light curves were constructed for each
of V9, V10 and V12, with the  sampling corresponding to the times of
real observations.  The amplitudes were selected to closely mimic the
light curves observed in Fig. 3. The fundamental period $P_0$ was not
detected  in power spectra of any of the simulated light curves. Next we
prewhitened our real and synthetic light curves by removing a sinusoid
with the main period. In all of the power spectra of the prewhitened
synthetic  light curves, the fundamental period and its aliases were
prominent. No periods close to $P_0$ were detected in any of the
prewhitened observed light curves, excluding the possibility of double
mode radial pulsations in these three variables.
 
The upper panel in Fig. 8 shows the power spectrum of  the observations
of variable V9 generated with CLEAN software (Roberts et al. 1987). The
most prominent peak is the radial frequency with period
$P_1=0.316307^d$. The arrow marks the position of the fundamental
period, assuming  $P_1/P_0$ = 0.745. A second, smaller peak is apparent
in the vicinity of the larger peak. To check on the reality of this
secondary peak we prewhitened our observed light curves removing a
sinusoid with the main period and its two harmonics. The power spectrum
of the prewhitened light curve of V9 is presented in the second panel
of Fig.  8. The highest peak now corresponds to the frequency of the
secondary peak in the first panel, with a period of $0.325128^d$. The
prewhitened observed light curve, phased with this period, is presented
in the lower panel of Fig. 8.
 
We performed a similar analysis for  variables V10 and V12. In the
power spectrum of V10 we again found two very close peaks (upper panel
of Fig. 9). After prewhitening of the observations, the highest peak in
the power spectrum corresponds to a period of $0.330363^d$ (middle
panel of Fig. 9). The lower panel of Fig. 9 shows the observations
phased with this period.  Variable V12 also shows multiperiodic behavior
(the upper panel of Fig. 10). After prewhitenning of the observations
the most dominant period is  $0.357818^d$ (the middle panel of Fig.
10).  The lower panel of Fig. 10 shows the observations phased with
this period.
 
Non-radial oscillations with many frequencies are common in
$\delta$ Scuti variables (main sequence stars in the instability
strip) but have not yet been observed in RR Lyr stars. The theoretical
calculations performed by Van Hoolst et al (1998) clearly show that
low-degree non-radial modes can be excited in RR Lyr stars. In their
model, a large number of unstable low-degree ($l=1,2$) modes in the
vicinity of the radial modes are partially trapped and therefore have
the largest growth rate and as a result are presumably most likely to be
excited  (see their Fig. 1). We propose that this is exactly what
we observe for variables V9, V10, and V12 in M55.
 
The $\delta$-Scuti stars have periods about 10 times shorter than RR
Lyr stars. Data sets consisting of observations over a few consecutive
nights cover many tens of  cycles of variability with good phase
coverage, with the result that the complex period structure can be well
determined. Observations of RR Lyr light curves are usually not so
extensive, with a typical light curve containing 100-200 measurements.
Our data are significantly more detailed in both phase coverage and
number of observations, with the light curves of our variables V9, V10
and V12 containing over 700 points collected during  four months. We
conclude that other RR Lyr variables may also pulsate with non-radial
modes but that one needs excellent  photometric coverage in order to
detect these non-radial frequencies.

\subsection{RRab variables}
 
Recently Jurcsik and Kovacs (1996), Kovacs and
Jurcsik (1996), Kovacs and Jurcsik (1997), and Jurcsik (1998) have 
extended the work of Clement, Simon and Teays, developing
methods for  obtaining the metallicity, absolute magnitudes, intrinsic
colors and temperatures of RRab stars basing on a Fourier
decomposition of $V$-band light curves. Their formulae are:
 
\begin{equation}
{\rm [Fe/H]} = -5.038 - 5.394\log P + 1.345\phi_{31}
\end{equation}
\begin{equation}
M_V = 1.221 - 1.396\log P - 0.477A_1 + 0.103\phi_{31}
\end{equation}
\begin{equation}
V_0-K_0=1.585 + 1.257\log P - 0.273A_1 - 0.234\phi_{31} + 0.062\phi_{41}
\end{equation}
\begin{equation}
\log T_{eff} = 3.9291 - 0.1112(V_0-K_0) - 0.0032{\rm [Fe/H]}
\end{equation}
 
\noindent where $\phi_{41}=\phi_4-4\phi_1$ (cf. equation 1).
 
The above equations are valid only for RRab stars with regular light
curves, i.e. variables with a deviation parameter $D_m$  smaller than 3
(see  Jurcsik and Kovacs 1996). In our sample, only variable V1
satisfies this condition. In order to increase the sample size, we
include  RRab variables with $D_m<5.5$, a condition satisfied by all of
the RRab stars in Table 1 belonging to M55. The results are listed in
Table 3 which contains values and errors of $A_0$, $A_1$, $\phi_{31}$,
$\phi_{41}$, $M_V$, [Fe/H], $T_{eff}$ and $D_m$.  The errors of $A_0$,
$A_1$, $\phi_{31}$ and $\phi_{41}$ come from the least squares fitting
(cf. equation 1) and the errors of $M_V$ and [Fe/H] are computed from
formulae given by Jurcsik and Kovacs (1996) and Kovacs and Jurcsik
(1996). 
 
In Fig. 11 we plot $<V>$ - $M_V$ versus $<V>$, where $<V>$ is the mean
$V$ magnitude for the variables. The horizontal line is  a fit to the
data, giving an apparent distance modulus of $13.65\pm0.11$, comparable
to the value of 13.76 listed by Harris (1996).
 
As discussed in Section 3.1, previous determinations of the metallicity
of M55 vary between $-1.54$ and $-1.89$. The Fourier  analysis of the
RRab stars is consistent with this range, with only variable V8
indicating a larger value of [Fe/H]. However, this star has the largest
value of  $D_m$ in the present sample of M55 RRab stars.  Our
determinations of [Fe/H] based on a Fourier decomposition of the light
curves of RRab variables are consistent with the period-amplitude
diagram presented in Fig. 5.
 
The reddening of RRab stars can be calculated from the metallicity,
expressed in terms of $\Delta S$ (Preston 1959). From Blanco (1992) we
have the relation 
\begin{equation} 
E(B-V) = <B-V>_{\Phi(0.5-0.8)} +0.01222\Delta S - 0.00045(\Delta S)^2 - 0.185 P - 0.356
\end{equation}
 
\noindent where $<B-V>_{\Phi(0.5-0.8)}$ is the observed mean color in
the 0.5--0.8 phase interval.

Based on the globular cluster metallicity scale adopted by Zinn
and West (1984) and Zinn (1985), Suntzeff et al. (1991) derived the
following $\Delta S$--[Fe/H] relation:
\begin{equation}
{\rm [Fe/H]} = -0.408 - 0.158 \Delta S
\end{equation}
 
Using above formulae and adopting a metallicity for M55 
of  ${\rm [Fe/H]}=-1.82$ (Zinn and West 1984) we obtain $\Delta S =
8.94$. 
 
The average value of $E(B-V)$ calculated in this way for our four M55
RRab stars is $E(B-V)=0.11\pm0.03$.  The reddening in Blanco's analysis
is fairly independent of the metallicity over the quoted range. Using a
metallicity of ${\rm [Fe/H]}=-1.54$ (Smith 1984) we derive
$E(B-V)=0.10\pm0.03$. A similar calculation for the reddening of the
RRab star in Sagittarius with a well defined light curve (V14 with $D_m$
= 1.40) gives $E(B-V)=0.09$, accepting the value of ${\rm [Fe/H]}$
determined from the Fourier analysis for this star.
 
These values are slightly larger than previous determinations of
between 0.06 and 0.08 mag (Lee 1977, Reed et al. 1988).  Recently
Schlegel et al. (1998) have published a new all-sky reddening map based
on the COBE/DIRBE and IRAS/ISSA maps. Their value of $E(B-V)$ at the
position of M55 $E(B-V)$ = 0.135 mag, consistent with our estimate.
 
Adopting a value of $E(B-V)=0.11\pm0.03$ we estimate the absolute
distance modulus  to be $(m-M)_0=13.31\pm0.11$.

\subsection{The instability strip}
 
The red and blue edges of the RR Lyrae zone on the horizontal branch
compiled by Smith (1995) vary in $(B-V)_0$ color between 0.155 and 0.19
mag for the blue edge and between 0.38 and 0.44 mag for the red edge. 
This corresponds to ranges in effective temperature of 7600--7400~K and
6250--6100~K, respectively. In our case the instability strip lies
between $B-V=0.31$ and $B-V=0.49$~mag. Using a color
excess of $E(B-V)=0.11$ mag, we obtain 
$(B-V)^{BE}_{0}=0.20$ and $(B-V)^{RE}_{0}=0.38$ mag for the edges of the
instability strip,
consistent within the errors with estimates made for
other globular clusters (Smith 1995). 
 
\section{Conclusions}
 
We have identified nine new RR Lyr variables in the field of the
globular cluster M55.  Two of these variables are probable members of
the Sagittarius dwarf galaxy.  The number of known RR Lyr variables in
M55 is now thirteen, four RRab stars and nine RRc stars.  The periods
of the variables indicate that M55 is an Oosterhoff type II cluster.
 
Three of our nine RRc variables exhibit marked amplitude modulation on
a time scale of less than a week. We excluded the hypothesis that such
behavior is caused by the Blazhko effect or double mode radial
pulsations.  A detailed analysis of the power spectra showed other
clear frequencies in the vicinity of the main peak. These frequencies
are too close to the main period to be radial pulsations. We conclude
that we have detected non-radial pulsation in these RRc stars.
 
We used Fourier decomposition of the $V$-band light curves of the  RR
Lyr variables to measure  luminosities, effective temperatures,
metallicities and masses of these variables. The mean mass of the RRc
stars in M55 is equal to $M=0.53\pm0.03 M_\odot$, the mean $\log
L=1.75\pm0.01$, and  the mean effective temperature
$T_{eff}=7193\pm27$~K. The helium abundance of M55 is $Y=0.27\pm0.01$.
The blue edge of the instability strip lays at $(B-V)_0\approx$0.20 mag
and the red edge at $(B-V)_0\approx$0.38 mag.

The values of ${\rm [Fe/H]}$ obtained from the Fourier decomposition of
the light curves of the RRc stars and the period-amplitude measurements
presented in Fig. 5 both indicate a metallicity of  $\sim -1.8 $ for
M55, consistent with the measurements of Suntzeff et al (1991),
McWilliam et al (1992), among others.
 
A similar analysis for the  RRab stars suggests a metallicity of  $\sim
-1.5$. This value is based only on one star with a deviation parameter
$D_m$ smaller than 3.0, and as a result is likely of low accuracy.

Using the Blanco (1992) dependence of the color excess $E(B-V)$ upon the
mean color at minimum light of RRab stars, the pulsation period of the
star, and spectroscopic parameter $\Delta S$ we obtained a color excess
for M55 of $E(B-V)=0.11\pm0.03$ for an adopted metallicity of  ${\rm
[Fe/H]}=-1.8$. We also estimated the values of the visual apparent and
dereddened distance moduli to be equal to $13.65\pm0.11$ and
$13.31\pm0.11$, respectively.

We identified in the M55 field two RRab variables belonging most
probably to the Sagittarius dwarf galaxy. From the analysis of light
curves we obtained for both of them ${\rm [Fe/H]}=-1.13$. This result is
consistent with earlier determinations of metallicity for RR Lyr
variables in the Sagittarius dwarf (Sarajednini and Layden 1995, Mateo
et al. 1995, Marconi et al. 1998).
 
\begin{acknowledgements} 
We would like to thank to Prof. W. Dziembowski,
Prof. B. Paczy\'nski and Dr. P. Moskalik for helpful hints and comments.
We are also grateful to Dr. G. Pojma\'nski for his software help.
AO, JK, WP and WK were supported by the Polish Committee of Scientific
Research through grant 2P03D-011-12 and by NSF grant AST-9528096 to
Bohdan Paczy\'nski. ASC acknowledges support by KBN grant 2P03C00112.
\end{acknowledgements}

\clearpage
 
\begin{table*}
\begin{center}
Table 1: Elements of the RR Lyrae variables in M55 \\
\vspace{1cm}
\begin{tabular}{rcccccccl}
\hline
\hline
Star & RA(2000) & Decl.(2000) & P[days] & $<V>$ & $<B>-<V>$ &  $A_V$ &
$A_1$ & Type \\
\hline
\hline
 V1 & 19:40:22.45 & -30:58:24.28 & 0.579978 & 14.38  & 0.36   & 1.28  &
0.438 & RRab\\
 V2 & 19:39:42.24 & -30:57:57.47 & 0.406147 & 14.41  & 0.39   & 0.44  &
0.228 & RRc \\
 V3 & 19:40:05.27 & -31:02:34.50 & 0.661987 & 14.28  & 0.40   & 0.85  &
0.358 & RRab\\
 V4 & 19:40:07.43 & -30:56:32.11 & 0.384164 & 14.33  & 0.39   & 0.40  &
0.203 & RRc \\
 V5 & 19:39:55.89 & -30:58:44.42 & 0.376146 & 14.32  & 0.37   & 0.43  &
0.223 & RRc \\
 V6 & 19:40:07.69 & -30:57:49.96 & 0.388821 & 14.38  & 0.39   & 0.47  &
0.234 & RRc \\
 V7 & 19:39:59.89 & -30:57:33.08 & 0.682573 & 14.26  & 0.46   & 1.03  &
0.346 & RRab\\
 V8 & 19:40:02.08 & -30:58:58.43 & 0.721961 & 14.37  & 0.49   & 0.62  &
0.248 & RRab\\
 V9 & 19:40:27.38 & -30:57:59.07 & 0.316307 & 14.43  & 0.31   & 0.41  &
0.198 & RRc \\
V10 & 19:40:08.51 & -30:54:49.04 & 0.331763 & 14.41  & 0.32   & 0.31  &
0.163 & RRc \\
V11 & 19:40:12.02 & -30:56:14.06 & 0.309954 & 14.42  & 0.31   & 0.21  &
0.105 & RRc \\
V12 & 19:39:59.80 & -30:58:02.73 & 0.325864 & 14.34  & 0.32   & 0.26  &
0.125 & RRc \\
V13 & 19:39:53.08 & -30:50:30.57 & 0.397841 & 14.44  & 0.34   & 0.39  &
0.195 & RRc \\
V14 & 19:39:54.76 & -30:50:10.17 & 0.521616 & 17.97  & 0.43   & 1.03  &
0.353 & RRab Sgr\\
V15 & 19:39:43.95 & -31:00:36.33 & 0.637286 & 18.33  & 0.42   & 0.40  &
0.164 & RRab Sgr\\
\hline
\hline
\end{tabular}
\end{center}
\end{table*}
 
\clearpage
 
\begin{table*}
\begin{center}
Table 2: Parameters for the RRc variables in M55.\\
\vspace{1cm}  
\begin{tabular}{rcccccr}
\hline
\hline
Star & $A_0$ & $A_1$ & $\phi_{31}$ & $M$ & $\log L$ & $T_{eff}$\\
 &
$\sigma_{A_0}$&$\sigma_{A_1}$&$\sigma_{\phi_{31}}$&$\sigma_M$&$\sigma_{\log
L}$&$\sigma_{T_{eff}}$ \\
\hline
\hline
 V2 & 14.418  & 0.228 & 4.089  & 0.545  & 1.766 & 7150\\
    &  0.000  & 0.001 & 0.039  & 0.005  & 0.002 &    3\\
\hline
 V4 & 14.342  & 0.203 & 4.216  & 0.513  & 1.733 & 7220\\
    &  0.001  & 0.001 & 0.049  & 0.006  & 0.003 &    4\\
\hline
 V5 & 14.335  & 0.223 & 3.947  & 0.543  & 1.739 & 7217\\
    &  0.000  & 0.001 & 0.042  & 0.006  & 0.002 &    3\\
\hline
 V6 & 14.363  & 0.234 & 3.847  & 0.567  & 1.760 & 7173\\
    &  0.001  & 0.001 & 0.074  & 0.011  & 0.004 &    6\\
\hline
V11 & 14.423  & 0.105 & 3.448  & 0.557  & 1.681 & 7375\\
    &  0.000  & 0.000 & 0.306  & 0.043  & 0.018 &   28\\
\hline
V13 & 14.445  & 0.195 & 4.448  & 0.493  & 1.736 & 7205\\
    &  0.001  & 0.001 & 0.059  & 0.007  & 0.003 &    5\\
\hline
\hline
\end{tabular}
\end{center}
\end{table*}
 
\clearpage
 
\begin{table*}
\begin{center}
Table 3: Parameters for the RRab variables in M55. \\
\vspace{1cm}
\begin{tabular}{rccccccrrr}
\hline
\hline
Star & P[days] & $A_0$ & $A_1$ & $\phi_{31}$ & $\phi_{41}$ & $M_V$ &
[Fe/H] & $T_{eff}$ & $D_m$ \\
 &
&$\sigma_{A_0}$&$\sigma_{A_1}$&$\sigma_{\phi_{31}}$&$\sigma_{\phi_{41}}$&$\sigma_{M_V}$&$\sigma_{[Fe/H]}$&
 & \\
\hline
\hline
 V1 & 0.579978 & 14.447  & 0.438  & 4.967  & 1.272  & 0.719 & -1.486 &
6461 & 2.62 \\
    &          &  0.000  & 0.001  & 0.005  & 0.007  & 0.084 &  0.022 &  
   &      \\
\hline
 V3 & 0.661987 & 14.316  & 0.358  & 5.013  & 1.494  & 0.647 & -1.867 &
6270 & 4.72 \\
    &          &  0.000  & 0.001  & 0.012  & 0.022  & 0.089 &  0.032 &  
   &      \\
\hline
 V7 & 0.682573 & 14.304  & 0.346  & 5.301  & 2.017  & 0.654 & -1.590 &
6267 & 4.44 \\
    &          &  0.001  & 0.001  & 0.008  & 0.013  & 0.094 &  0.034 &  
   &      \\
\hline
 V8 & 0.721961 & 14.383  & 0.248  & 5.663  & 2.271  & 0.684 & -1.316 &
6243 & 5.12 \\
    &          &  0.001  & 0.001  & 0.016  & 0.038  & 0.102 &  0.047 &  
   &      \\
\hline
\hline
V14 & 0.521616 & 18.014  & 0.353  & 4.996  & 1.457  & 0.844 & -1.132 &
6520 & 1.40 \\
    &          &  0.001  & 0.001  & 0.011  & 0.017  & 0.085 &  0.024 &  
   &      \\
\hline
V15 & 0.637286 & 18.341  & 0.164  & 5.457  & 2.602  & 0.821 & -1.135 &
6258 & 18.46 \\
    &          &  0.001  & 0.001  & 0.045  & 0.093  & 0.097 &  0.067 &  
   &       \\
\hline
\hline
\end{tabular}
\end{center}
\end{table*}
 
\clearpage
 
\begin{figure}
\plotone{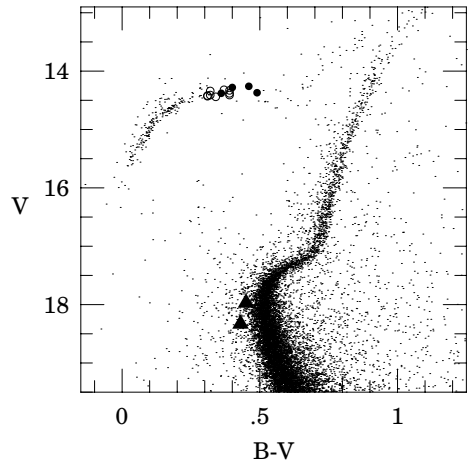}
\caption{A color-magnitude diagram of M55. The filled circles, open
circles and triangles denote RRab and RRc stars from M55 and RRab
variables from the Sagittarius dwarf galaxy, respectively.}
\end{figure}
 
\begin{figure}
\plotone{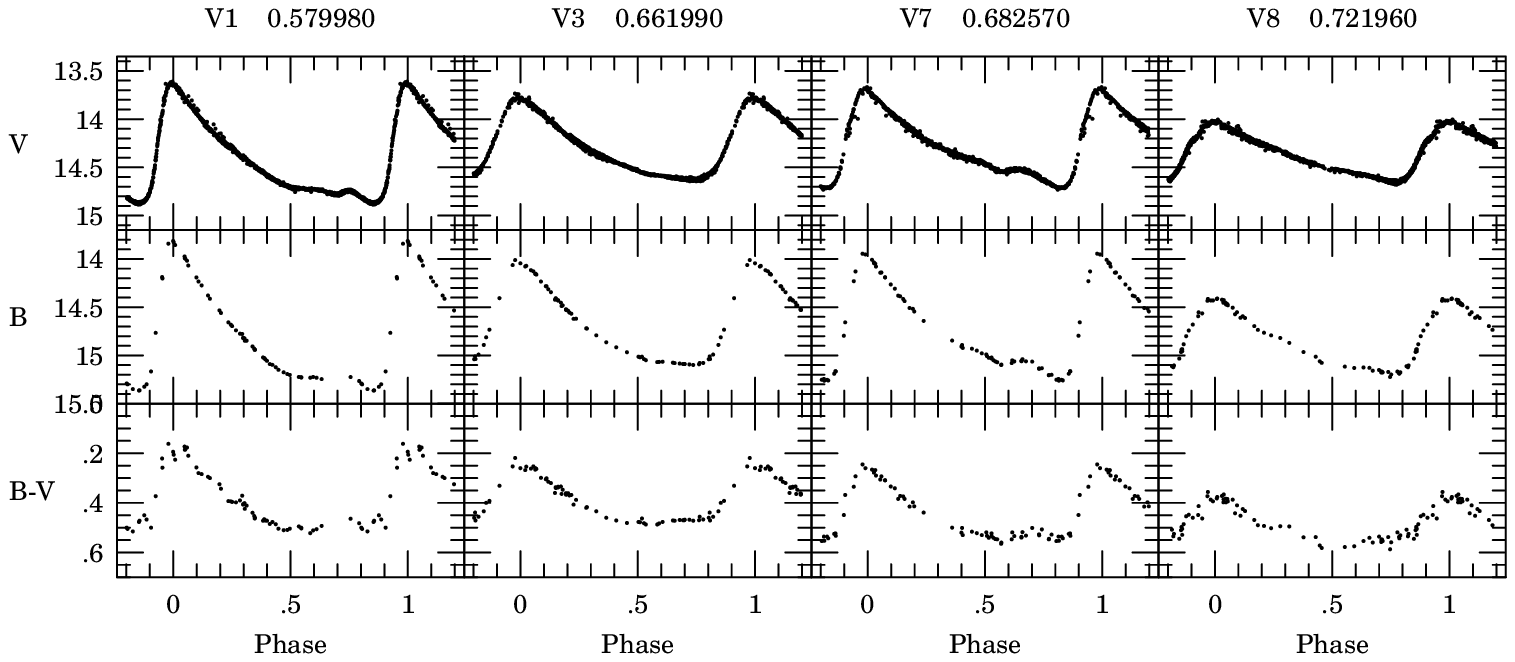}
\caption{ The $V$, $B$ and $B-V$ light curves of RRab Lyr
stars from M55.}
\end{figure}
 
\begin{figure}
\plotone{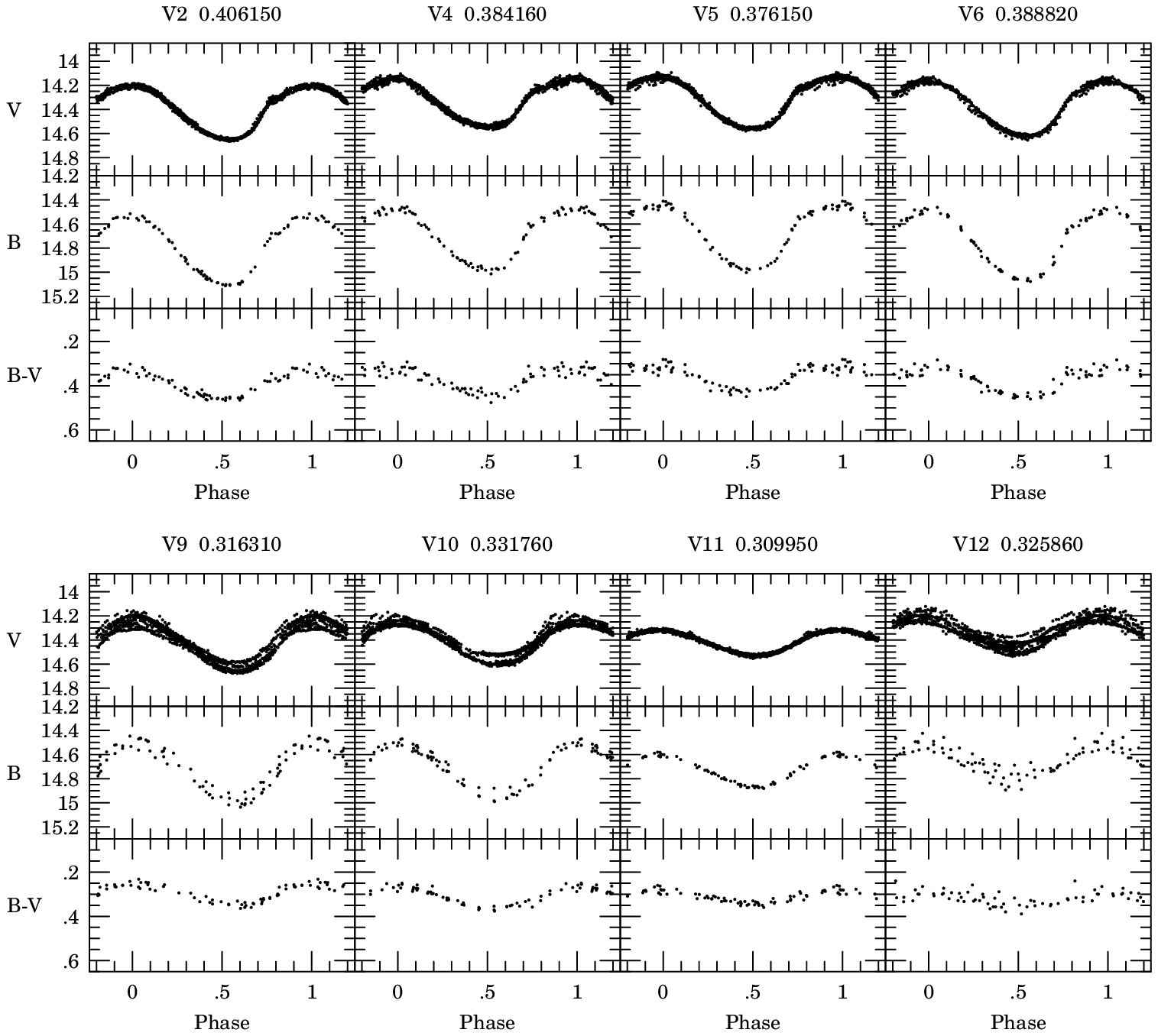}
\caption{ The $V$, $B$ and $B-V$ light curves of RRc Lyr
stars from M55.}
\end{figure}
 
\begin{figure}
\plotone{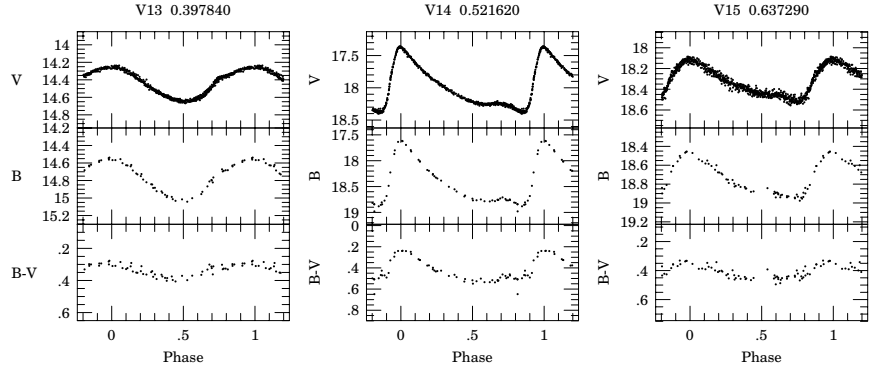}
\caption{The $V$, $B$ and $B-V$ light curves of RRc Lyr
star from M55 and two RRab Lyr stars from Sagittarius Dwarf Galaxy.}
\end{figure}

\begin{figure} 
\plotone{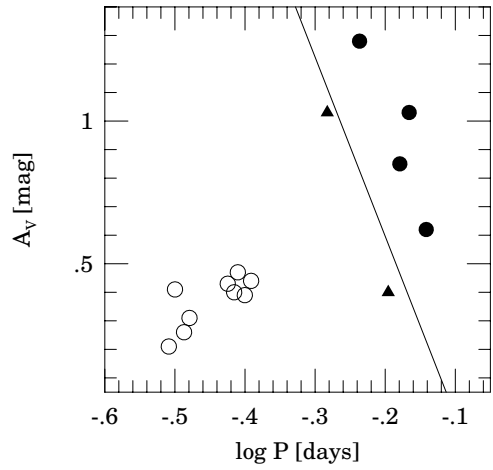} 
\caption{The
period-amplitude diagram form RRab stars from M55 (filled    circles),
RRc stars from M55 (open circles) and RRab stars from Sagittarius Dwarf
Galaxy (triangles). The solid line represents a linear fit to RRab
variables in M3 (Kaluzny et al 1998).} 
\end{figure}
 
\begin{figure}
\plotone{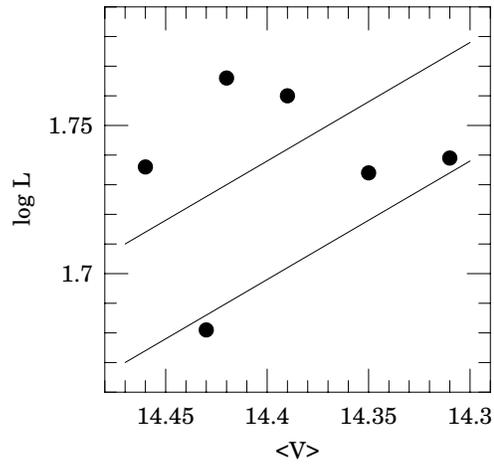}
\caption{The dependence between luminosity and visual magnitude for
the RRc stars in  M55. The solid lines
have a slope of 0.4 and are separated by 0.04 in $\log L$, which
represents the uncertainty in the values of $\log L$ computed from
$\phi_{31}$ and $P$.}
\end{figure}
 
\begin{figure}
\plotone{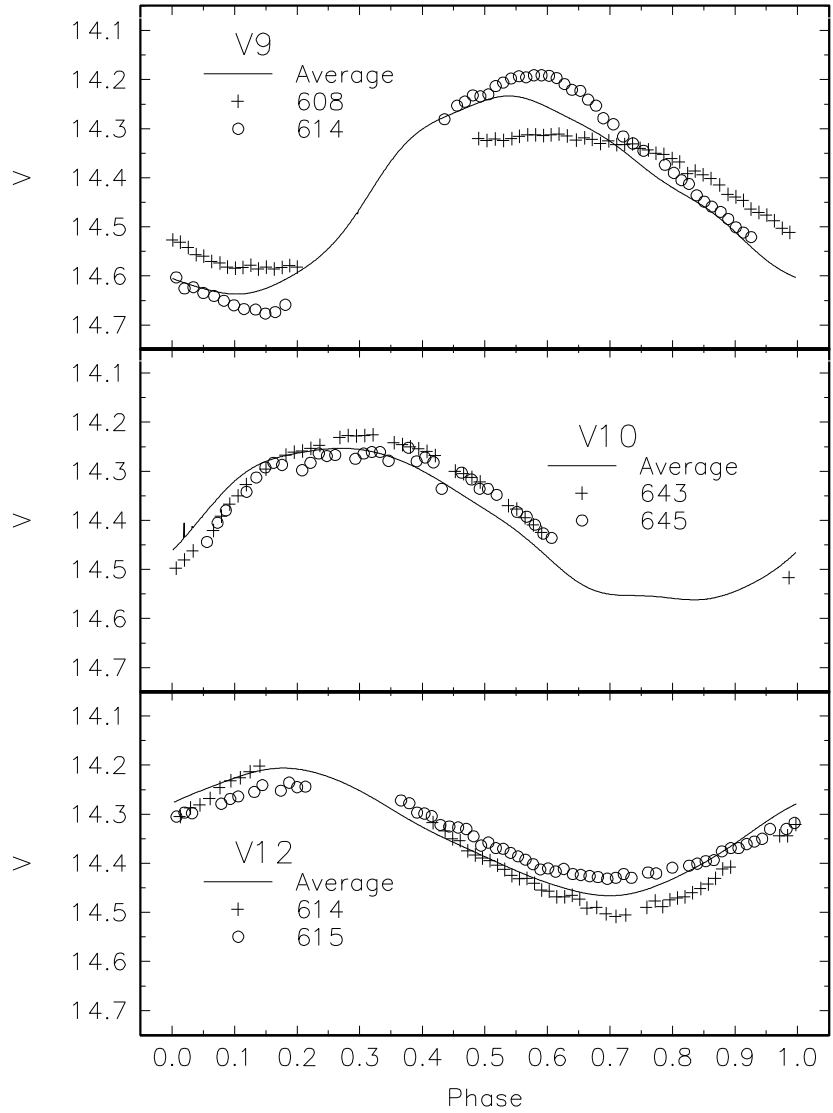}
\caption{The irregular light curves of three RRc stars. Light curves for
individual nights are labeled with 
truncated HJD numbers.}
\end{figure}
 
\begin{figure}
\plotone{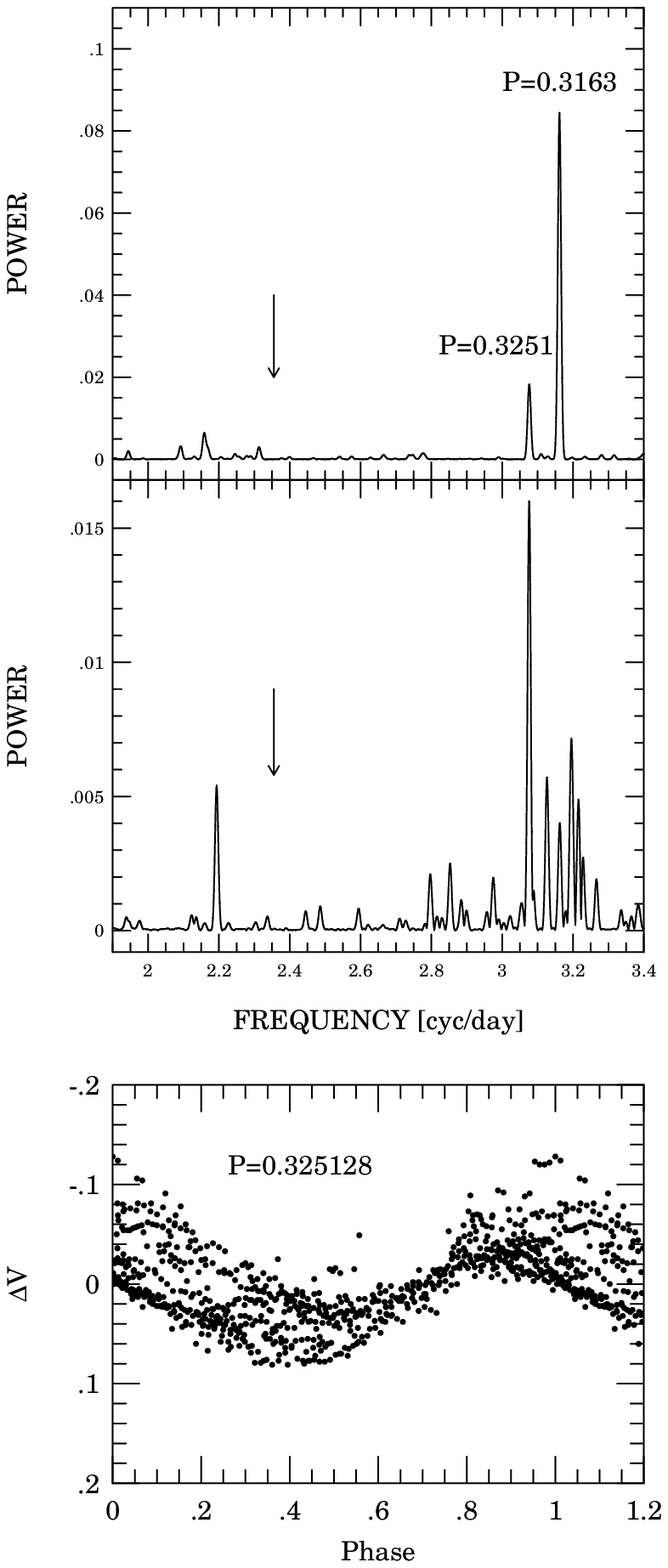}
\caption{The power spectra of the real and prewhitened light curve of
V9 (upper and middle panel). The arrow marks the suspected position of
the fundamental period. The lower panel shows the light curve of the
star
phased with the period corresponding to the highest peak in the middle
panel.}
\end{figure}
 
\begin{figure}
\plotone{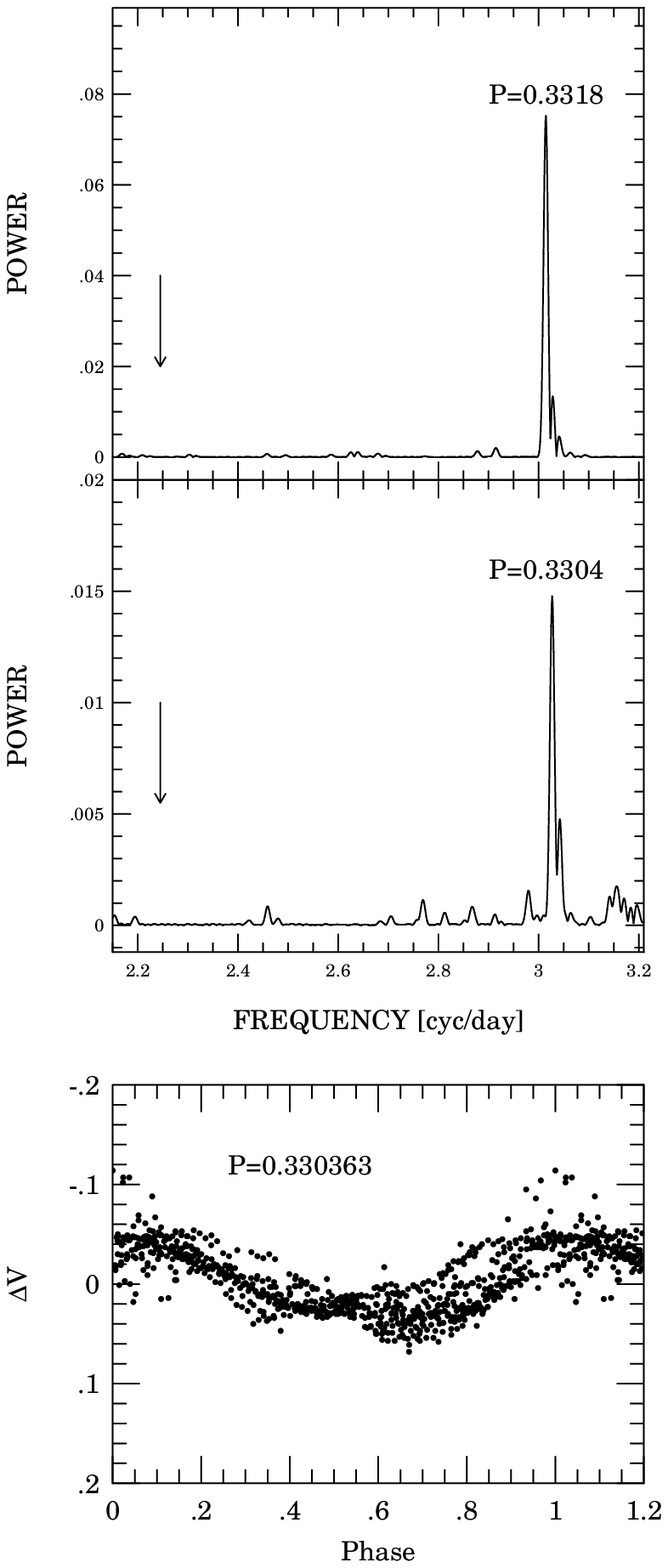}
\caption{The power spectra of the real and prewhitened light curve of
V10 (upper and middle panel). The arrow marks the suspected position of
the fundamental period. The lower panel shows the light curve of the
star phased with the period corresponding to the highest peak in the
middle
panel.}
\end{figure}
 
\begin{figure}
\plotone{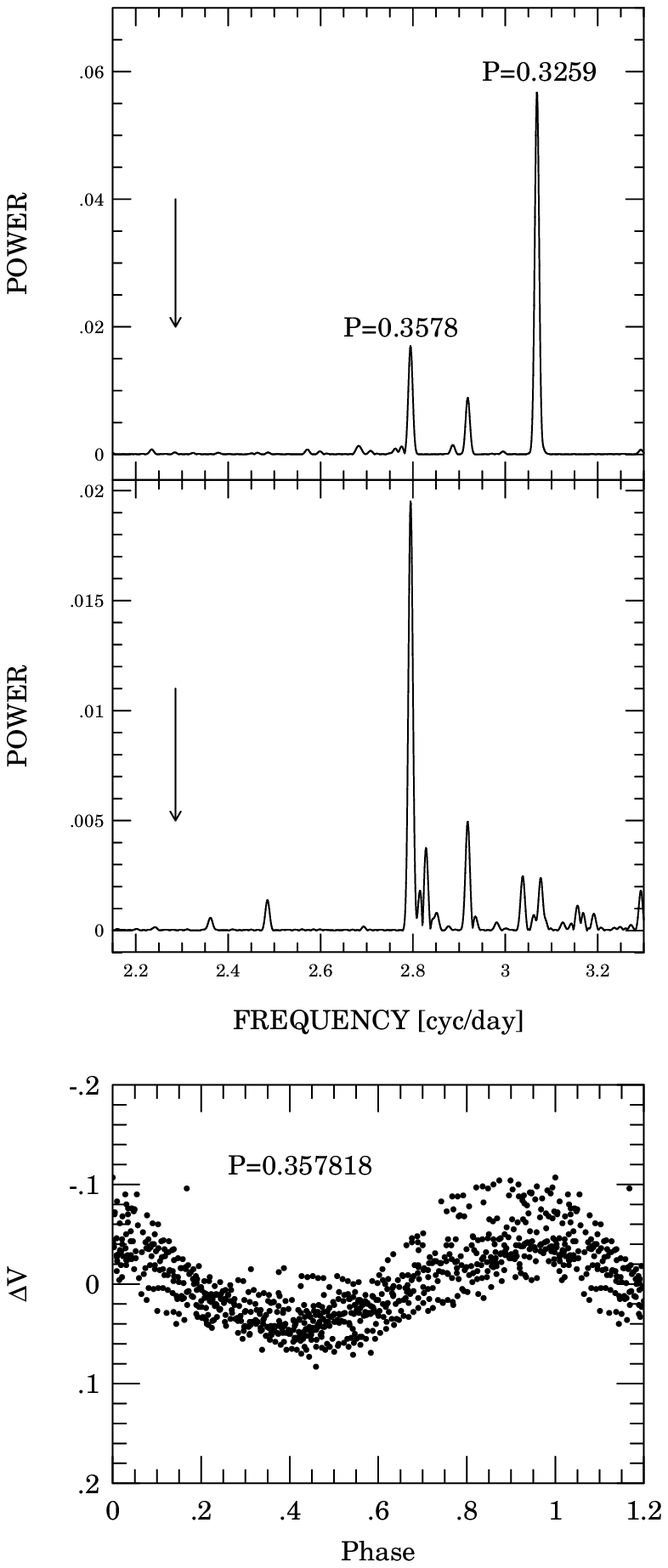}
\caption{The power spectra of the real and prewhitened light curve of
V12 (upper and middle panel). The arrow marks the suspected position of
the fundamental period. The lower panel shows the light curve of the
star phased with the period corresponding to the highest peak in the
middle
panel.}
\end{figure}
 
\begin{figure}
\plotone{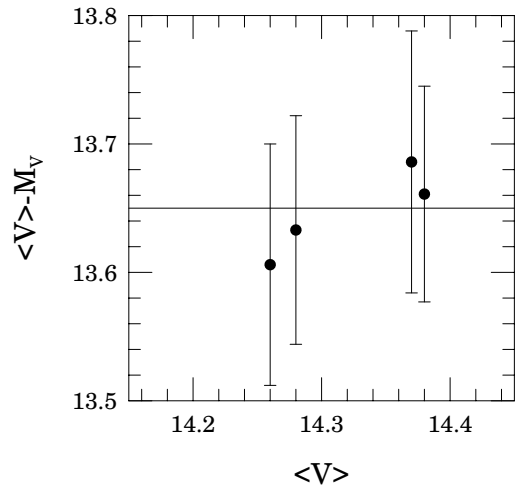}
\caption{The $<V>-M_V$ versus $<V>$ for the M55 RRab variables.
The horizontal line is a fit to the data, giving an apparent distance
modulus of $13.65\pm0.11$.}
\end{figure}

\end{document}